\documentclass[twocolumn]{revtex4}

\usepackage{graphicx}
\usepackage{epsfig}

\begin{document}

\title{The cosmological dark sector as a scalar $\sigma$-meson field}

\author{Saulo Carneiro}

\affiliation{Instituto de F\'{\i}sica, Universidade Federal da Bahia, 40210-340, Salvador, BA, Brazil}

\begin{abstract}
Previous quantum field estimations of the QCD vacuum in the expanding space-time lead to a dark energy component scaling linearly with the Hubble parameter, which gives the correct figure for the observed cosmological term. Here we show that this behaviour also appears at the classical level, as a result of the chiral symmetry breaking in a low energy, effective $\sigma$-model. The dark sector is described in a unified way by the $\sigma$ condensate and its fluctuations, giving rise to a decaying dark energy and a homogeneous creation of non-relativistic dark particles. The creation rate and the future asymptotic de Sitter horizon are both determined by the $\sigma$ mass scale.
\end{abstract}

\maketitle


The cosmological constant problem is often stated as a large, non-observed contribution of the vacuum of quantum fields to the dark energy component of the cosmic fluid. Although very popular, this statement turns the problem ill defined, because the quantum field calculations are usually performed in the flat space-time, where nothing would really gravitate. A more meaningful procedure is to obtain the vacuum contribution in the curved, FLRW space-time, with a reliable renormalisation that absorbs the divergent flat result in the constants of the theory \cite{conformal}. This can be made, for example, in the case of conformal fields, which naturally leads to a vacuum density of the order of $H^4$ ($H$ is the Hubble function), since no other scale is available. In the present universe, this figure has the opposite problem of being too small as compared to the observed cosmological term. More generally, other even powers of $H$ can result from these calculations, owing to a theorem on the allowed counter terms in the effective action \cite{Wald}. For instance, a free field of mass $m$ may lead to a contribution of order $m^2 H^2$, which has the observed figure if $m$ has the order of the Planck mass. However, with a dark energy scaling with $H^2$, the cosmic expansion would not experience the observed transition from decelerating to accelerating phases. The remaining possibility is a constant $\Lambda$, which value, however, is not fixed by the theory.

The theorem mentioned above has, as always, its premises. Among them we can find the conditions that the fields are free, and that the vacuum energy-momentum tensor is conserved. The first does not apply, for example, if we estimate the vacuum contribution of the low energy, strongly coupled fields of QCD. Indeed, some studies have suggested that these fields lead to a vacuum term scaling as $m_{\pi}^3 H$, where $m_{\pi}$ is the energy scale of the chiral phase transition \cite{Schutzhold}. Surprisingly enough, this gives the observed order of magnitude for $\Lambda$. The second premise, as well, cannot be always satisfied. A decaying $\Lambda$ is only possible either if its equation of state parameter $\omega \neq -1$, or if it interacts with dark matter. As shown elsewhere \cite{CB,wands,winfried}, $\omega = -1$ is a necessary condition for $\Lambda$ to be strictly homogeneous, which permits an unambiguous definition of clustering dark matter. Therefore, a vacuum term decaying with $H$ is necessarily concomitant with a homogeneous creation of dark particles. The freedom in defining $\Lambda$ such that $\omega = -1$ is a manifestation of the so called dark degeneracy \cite{CB,Degeneracy}. As we will see, this freedom allows a unified description of the dark sector in terms of a non-adiabatic scalar field, which plays the role of both dark energy and dark matter. In particular, we will show that the $\sigma$ scalar, that emerges from effective models of low energy QCD, produces indeed a dark energy scaling linearly with $H$.

In its simplest version, the $\sigma$ potential is given by \cite{tHooft}
\begin{equation}\label{1}
V = - \frac{\mu^2}{2} \sigma^2 + \frac{\lambda}{8} \sigma^4 + \tilde{V},
\end{equation}
where $\mu$, $\lambda$ and $\tilde{V}$ are positive constants.
It presents a false vacuum at $\sigma = 0$ and a true vacuum at $\sigma = f$, with $f$ given by
$\lambda f^2 = 2 \mu^2$.
There is a priori no clear prescription for the choice of vacuum in cosmological backgrounds. Nevertheless, as our universe is approaching a de Sitter space-time with very small curvature, we assume that the flat space-time vacuum structure is not altered, except for the minimum of the potential at the true vacuum state, as we will show below.
When the chiral symmetry is spontaneously broken, expanding the potential (\ref{1}) around the vacuum expectation value $f$ leads, up to second order, to
\begin{equation}\label{4}
V(\phi) = V_{dS} + \frac{M^2 \phi^2}{2},
\end{equation}
where $\phi = \sigma - f$ are scalar fluctuations of the chiral condensate, with mass $M = \sqrt{2}\mu$, and
\begin{equation}\label{3}
V_{dS} = \tilde{V} - \frac{\mu^2 f^2}{4}.
\end{equation}
We will see that, in an FLRW background, $V_{dS}$ is also determined by $\mu$.
The energy density and pressure of the scalar field are
\begin{eqnarray}\label{5a}
\rho_{\phi} &=& \frac{\dot{\phi}^2}{2} + V, \\\label{5b}
p_{\phi} &=& \frac{\dot{\phi}^2}{2} - V,
\end{eqnarray}
where a dot means derivative w.r.t. the cosmological time. We can always decompose this perfect fluid into pressureless matter and a $\Lambda$ component \cite{CB,Degeneracy}, such that
\begin{eqnarray}\label{6a}
\rho_m &=& \dot{\phi}^2,\\\label{6b}
\Lambda = - \,p_{\Lambda} &=& V - \frac{\dot{\phi}^2}{2}.
\end{eqnarray}
In the general case, these two components interact and, in this way, the content above can be seen as a particular case of a non-adiabatic dark sector \cite{iModels,Pigozzo,scalar}.
This decomposition assures that the $\Lambda$ component does not cluster, provided the matter component follows geodesics, i.e. provided there is no momentum transfer between the components \cite{CB,wands,winfried}. We are going to show that this is indeed the case. Furthermore, as $\delta \Lambda = 0$, the scalar-field effective sound speed $\delta p_{\phi}/\delta \rho_{\phi}$ is zero, which prevents, at the perturbation level, oscillations and instabilities in the matter power spectrum \cite{CB}. 

The energy-momentum balance between the two components can be written as
\begin{eqnarray}\label{balance1}
T_{m;\nu}^{\mu\nu} &=& Q^{\mu},\\ \label{balance2}
T_{\Lambda;\nu}^{\mu\nu} &=& -Q^{\mu},
\end{eqnarray}
where
\begin{equation}\label{Q}
Q^{\mu} = Q u^{\mu} + \bar{Q}^{\mu} \quad \quad (\bar{Q}^{\mu} u_{\mu} = 0)
\end{equation}
is the energy-momentum transfer, and $u^{\mu}$ is the fluid $4$-velocity. The transverse component $\bar{Q}^{\mu}$ represents the momentum transfer. Writing for the $\Lambda$ term
\begin{equation}
T_{\Lambda}^{\mu\nu}  = \Lambda g^{\mu\nu},
\end{equation}
we have, from (\ref{balance2}),
\begin{eqnarray}
Q &=& - \Lambda_{,\nu} u^{\nu},\\
\bar{Q}^{\mu} &=& \Lambda_{,\nu} (u^{\mu} u^{\nu} - g^{\mu\nu}).
\end{eqnarray}
Perturbing these equations we obtain $\delta \bar{Q}_0 = 0$ and, in a co-moving gauge,
\begin{eqnarray}
\delta Q &=& -(\delta \Lambda)_{,0} \, ,\\
\delta \bar{Q}_i &=& (\delta \Lambda)_{,i}.
\end{eqnarray}
Therefore, if we show that $\delta \bar{Q}^{\mu}$ is zero, we show that $\Lambda$ is strictly homogeneous (and, in addition, that $\delta Q = 0$).
Components (\ref{6a}) and (\ref{6b}) can be written, respectively, in the covariant form
\begin{eqnarray}\label{matter}
T_m^{\mu\nu} &=& \partial^{\mu}\phi \, \partial^{\nu}\phi, \\ \label{lambda}
T_{\Lambda}^{\mu\nu}  &=& \left( V - \frac{1}{2} \partial_{\alpha}\phi \, \partial^{\alpha}\phi \right) g^{\mu\nu}.
\end{eqnarray}
From (\ref{lambda}) and (\ref{balance2}) we can write
\begin{equation}
Q_{\mu} = \partial_{\alpha}\phi \, \partial^{\alpha}\partial_{\mu}\phi - V' \,\partial_{\mu}\phi,
\end{equation}
with $V' = M^2 \phi$ from (\ref{4}). In Fourier's space it takes the form
\begin{equation}
Q_{\mu} = ik_{\mu} (M^2 - k^2) \phi^2,
\end{equation}
where $k_{\mu}$ is the scalar-field wavevector. By doing $i k_{\mu} = k u_{\mu}$ we obtain
\begin{equation}
Q_{\mu} = k (M^2 - k^2) \phi^2 u_{\mu},
\end{equation}
which shows that $\bar{Q}^{\mu} = 0$ in any space-time \footnote{Technically, any space-time admiting plane waves as a complete ortogonal basis, in particular the spatially flat FLRW space-time and its first order perturbations.}.

The Friedmann equations for the spatially flat FLRW space-time are given by
\begin{eqnarray}\label{7a}
3H^2 &=& V + 2H'^2,\\\label{7b}
\dot{\phi} &=& - 2 H',
\end{eqnarray}
where a prime means derivative w.r.t. the scalar field. On the other hand, the Klein-Gordon equation,
\begin{equation}\label{8}
\ddot{\phi} + 3H\dot{\phi} + V'= 0,
\end{equation}
can be put in the form
\begin{equation}\label{9}
\dot{\rho}_m + 3 H \rho_m = - \dot{\Lambda} \equiv \Gamma \rho_m.
\end{equation}
The last equality defines the rate of matter creation $\Gamma$. Using (\ref{6a}), (\ref{6b}), (\ref{7b}) and (\ref{8}) into (\ref{9}), we obtain
\begin{equation}\label{10}
V'= (3H + \Gamma) H'.
\end{equation}
With the potential given by (\ref{4}), equations (\ref{7a}) and (\ref{10}) determine the solutions for $H(\phi)$ and $\Gamma(\phi)$.
The second order solution for $H$ can be written as
\begin{equation} \label{11}
H = H_{dS} \, (1 + A \phi^2).
\end{equation}
Using (\ref{4}) and (\ref{11}) into (\ref{10}) we obtain, at the same order of approximation,
\begin{equation}\label{12}
\Gamma = \frac{M^2}{2AH_{dS}} - 3H_{dS},
\end{equation}
which means that matter is created at a constant rate. On the other hand, from (\ref{6a}), (\ref{7b}) and (\ref{9}) we have
\begin{equation}
\Lambda'= 2\Gamma H',
\end{equation}
which, after integration, leads to
\begin{equation}
\Lambda = 2\Gamma H + \tilde{\Lambda},
\end{equation}
where $\tilde{\Lambda}$ is an arbitrary integration constant. There is no energy scale fixing $\tilde{\Lambda}$, let us hence take it zero, that is,
\begin{equation}\label{13}
\Lambda = 2 \Gamma H.
\end{equation}
In the de Sitter limit $\phi \rightarrow 0$ we have, from (\ref{4}), (\ref{7a}), (\ref{11}) and (\ref{13}),
\begin{equation}\label{15}
V_{dS} = 3H_{dS}^2 = \frac{4\Gamma^2}{3}.
\end{equation}
Substituting (\ref{11}) into (\ref{7a}), retaining only terms up to second order in $\phi$ and using (\ref{15}) we obtain
\begin{equation} \label{16}
\frac{32\Gamma^2}{9} A^2 - \frac{8\Gamma^2}{3} A + \frac{M^2}{2} = 0.
\end{equation}
On the other hand, using (\ref{15}) into (\ref{12}) we get
\begin{equation}\label{17}
M^2 = 4 \Gamma^2 A.
\end{equation}
From (\ref{16}) and (\ref{17}) we then have $A = 3/16$, which, substituted back into (\ref{17}), gives
\begin{equation}\label{19}
\Gamma^2 = \frac{4M^2}{3}.
\end{equation}
We see that $\sqrt{2}\mu = M$ completely determines not only the mass of $\phi$, but, through (\ref{15}) and (\ref{19}), also the matter creation rate $\Gamma$, the de Sitter horizon $H_{dS}$ and the potential (\ref{4}).
The cosmological solution with constant-rate creation of matter is equivalent to a non-adiabatic generalised Chaplygin gas with $\alpha = -1/2$ \cite{Pigozzo} and was tested against observations at background and perturbation levels, showing a good approximation to the present universe \cite{PLB}. For $\Gamma = 0$ we have a Minkowski space-time with $f = 0$ and $V=0$, which is however unstable and eventually collapses to a singularity \cite{dynamical}. For $\Gamma \neq 0$ we have an expanding space-time, which tends asymptotically to a stable de Sitter universe. 

As a matter of fact, the result (\ref{13}) can be achieved with any free scalar field of mass $M$, since all we need is a potential like (\ref{4}) and a space-time close enough to de Sitter \cite{CB}. Nevertheless, we may show that the QCD chiral phase transition leads naturally to the correct energy scale we need to describe the observed universe.
From (\ref{13}) we see that $ \Gamma$ has the dimension of the cube of a characteristic energy $m_{\pi}$. By taking (\ref{13}) at the present time, we can write
\begin{equation}
m_{\pi}^3 \sim H_0 \Omega_{\Lambda0},
\end{equation} 
where $\Omega_{\Lambda0} \sim 1$ is the dark energy density relative to the critical density. By taking also $H_0 \sim 10^{-18} s^{-1}$, we have
\begin{equation}\label{28}
m_{\pi} \sim 100 \, {\rm MeV},
\end{equation}
which is the energy scale of the QCD phase transition. This can be heuristically understood if we estimate the energy of vacuum fluctuations in a de Sitter background as $\delta E \sim H$. In the case of a free massless field, the vacuum density is then given by $\delta E \, H^{3} \sim H^4$, a result that, in the high energy limit, is useful for obtaining non-singular cosmological solutions \cite{chimento}. However, in the case of a low energy, strongly interacting field, fluctuations are confined to a volume $m_{\pi}^{-3}$, where $m_{\pi}$ is the energy scale of confinement. Hence, the vacuum density should be given, instead, by $m_{\pi}^3 H$.
Leading $m_{\pi}^3 \sim \Gamma$ into (\ref{19}), we obtain the mass \footnote{The cosmological role of a pseudo Nambu-Goldstone boson with mass $~ H_0$ was already considered before, in other contexts \cite{Waga}.}
\begin{equation}\label{29}
M \sim 10^{-33} \, {\rm eV}.
\end{equation}
On the other hand, if we assume that the above proportionality between the mass of the chiral condensate fluctuations and the condensate density is also valid at nuclear scales, we can write
\begin{equation}
\frac{M}{\rho_c} \sim \frac{m_{\sigma}}{\rho_{Nuc}} \sim m_\pi^{-3},
\end{equation}
\newline
where $\rho_c$ is the cosmological critical density, $m_{\sigma} \sim 1$ GeV is the observed $\sigma$ mass, and $\rho_{Nuc}$ is a typical nuclear density. With $\rho_c \sim 10^{-26}$ kg/m$^3$ and $\rho_{Nuc} \sim 10^{16}$ kg/m$^3$, we obtain for $M$ the same figure as in (\ref{29}). 
It is also worthy of note that $M$ is the quantum of energy expected in the de Sitter space-time. Indeed, the number of degrees of freedom inside the de Sitter horizon is given by the holographic bound $N \sim H_{dS}^{-2}$ \cite{Bousso,Guillermo}. The corresponding energy is $E \sim \rho \,H_{dS}^{-3} \sim H_{dS}^{-1}$. Therefore, the energy per degree of freedom is $E/N \sim H_{dS} \sim M$, by (\ref{15}) and (\ref{19}). It is also suggestive (see (\ref{15})) that the rate of dark particles creation $\Gamma$ has the order of the de Sitter horizon temperature $H_{dS}$ \cite{gibbons}.

\section*{Acknowledgements}

I am thankful to M. R. Robilotta, C. Chirenti, H. A. Borges, T. S. Pereira and W. Zimdahl for helpful discussions, to G. A. Mena Marug\'an, J. S. Alcaniz and J. C. Fabris for critical readings, and to CNPq for financial support.

\end{document}